\documentstyle[epsfig]{article}

\begin{document}

\date{}

\title{Fine Structure Constants in n-dimensional Physical
Spaces through Dimensional Analysis}

\author{Fabr\'{\i}cio Casarejos, Jaime F. Villas da Rocha
 \\ and Roberto Moreira Xavier \\
\small{\it{Instituto de F\'{\i}sica,}} 
Universidade do Estado do Rio de Janeiro, \\ \small
R. S. Francisco Xavier $524$, Maracan\~ a, $20550-013$ 
 Rio de Janeiro~--~RJ, Brazil\\ \small e-mail adresses:
 fabll@dft.if.uerj.br, roch@dft.if.uerj.br}
\maketitle

\begin{abstract}
We use Vaschy-Buckhingham Theorem  as a systematic tool
to build univocal n-dimensional extensions 
of the electric and gravitational fine structure
constants and show that their ratio is dimensionally invariant.
The results allow us to obtain the relative standard uncertainty
on the three-dimensionality of space as 1.08 $10^{-13}$.


\end{abstract}

Dimensional Analysis can be considered a cornerstone of the syntaxis
of Physics. Its axiomatic foundations, now well established, are based 
on the structure of a finite multiplicative group. A fundamental
result is the $\Pi$ or Vaschy-Buckingham Theorem
\cite{buc,Vaschy92} from which
we can infer the existence of a definite number of $\Pi$s, i. e.,
independent adimensional quantities, 
associated to the physical processes under consideration, 
 expressed as products of dimensional ones.
Fundamental physical theories
- Electrodynamics, Quantum Mechanics and General 
Relativity - involve fundamental
physical constants: $e$, the 
electron charge, $\hbar$, the Planck constant,  $c$,
the velocity of light and $G$, the newtonian gravitational constant.
In many applications of those theories, these constants appear in adimensional
multiplicative combinations, such as $\alpha$ $=$ $e^2/\hbar c$,
the fine structure constant for electromagnetic interactions
introduced by Sommerfeld
when he extended Bohr theory to include elliptical orbits and 
relativistic effects in order to explain the observed splitting of the
energy levels of hydrogen atoms in four-dimensional
spacetimes, i.e., in spaces with three dimensions.
Its intriguing numerical value, $\sim$  $1/137$, has motivated 
the until now vain search for its deduction from fundamental principles.
The importance of $\alpha$ has inspired also the construction 
of analogous fine structure constants involving other interactions
\cite{CarreRees}. 
We shall now discuss the dimensionality of space itself. 
The connection between the dimensionality 
of physical space and the mathematical structure of 
physical laws was first established by Ehrenfest \cite{ehr,ehr1}
who solved the n-dimensional Kepler problem 
showing that stable solutions are possible only for
n $=$ 2 or n $=$ 3. He treated also the 
stability of the Bohr model of the atom. 
This approach was later explored by other 
authors \cite{tanger,gure}. An alternative procedure to the
stability arguments has been presented 
\cite{caruso}. The idea of using 
nointeger dimensionalities and measuring $\epsilon$, 
the deviation from the usual three-dimensionality, according 
to Jammer \cite{Jam89} was first proposed
by Zeilinger and Svozil \cite{SZ85}
who analyzed the experimentally measured values
of the anomalous magnetic moment of the electron.
It has also been suggested that 
 M\"ossbauer effect
could be used to measure 
$\epsilon$, in a Pound-Rebka
device, by using the gravitational
redshift 
\cite{moreira}.
Now, to define our problem we introduce and interpret $\Pi$ numbers 
associated to interactions in n-dimensional spaces, using
the $M, L, T$ basis.

By Vaschy-Buckingham Theorem, given 
$i$ physical quantities $a_i$, with $k$ independent dimensions,
we can define $i$ $-$ $k$  adimensional $\Pi$
numbers \cite{baren}
\begin{equation}
\label{pi}
\Pi_j \; = \;  \frac{a_{k+j}}{a_1^{p_{k+j}}...a_k^{r_{k+j}}}
.
\end{equation}
\noindent This allows us to elect each $a_{k+j}$ as a directive
quantity according to the physical process under study. Directive 
quantities physically guide the construction of $\Pi$ numbers and 
correspond to Buckingham's derived quantities \cite{buc}. 
In n-dimensional physical spaces, 
$c$, a ratio
between space and time, and $\hbar$, related to unidimensional oscillators
(the only possible unidimensional structure),
maintain their dimensional formulas $LT^{-1}$ and
$ML^2T^{-1}$, respectively. From a similarity 
principle\footnote{ The similarity principle itself, which plays a 
fundamental role on these results, will be discussed in detail elsewhere.},
we obtain the n-dimensional Laplace Equation whose solution
yields interactions proportional to $1/L^{(n-1)}$. Then, if 
we preserve, also by a similarity principle, Newton's second law 
as second derivative of displacement, 
by the Dimensional Homogeneity Principle,
a straightforward calculation shows that
the 
n-dimensional newtonian gravitational and electromagnetic constants,
$G_n$ and $e_n$, 
have as dimensional formulas $L^n{M^{-1}T^{-2}}$ and 
$M L^n{T^{-2}}$.    

Let us consider now $e_n$, $\hbar$, $c$, $G_n$ and, 
as the gravitational 
``charge'', the 
proton mass $m_p$.
As we have a three-dimensional basis and five quantities, 
by the Vaschy-Buckingham Theorem,
we can construct only two independent $\Pi$ numbers. 
Interactions
are always given by the product of their characteristic charges.
This leads us necessarily to choose ${e_n}^2$ and $m_p^2$ as the
directive quantities. Then, Eq. (\ref{pi}) gives us
\begin{equation}
\label{alfa3}
       {\Pi}_{q_{{}}} = [\hbar]^{\epsilon_1}\;   \; 
       [{G_n}]^{\epsilon_2} \; \; 
[c]^{\epsilon_3}\;  \; {q}^2  ,
\end{equation}  where $q$ represents the charges.
The solutions of the respective systems are 
\begin{eqnarray}
\label{alfa4}
       {\Pi}_{{}_{{e_n}}} & = & \left[ {\hbar}^{2(2-n)} \;  \; 
        {G_n}^{{3-n}}
        \; \;
c^{2(n-4)}\right]^{\frac{1}{n-1}} \;  \; {e_n}^2 , \\
\label{alfa7}
       {\Pi}_{{{}_{m_p}}} & = & \left[{\hbar}^{{(2-n)}} 
       \; 
        \; 
       {G_n} 
\;  \; c^{{(n-4)}}\right]^{\frac{2}{n-1}} \;  \; {m_p}^2  .
\end{eqnarray}  

Without the Vaschy-Buckingham Theorem, 
we could not have obtained these univocal solutions. The
remarkable singularity at $n$ $=$ $1$ 
appears also in the n-dimensional extension
of the Planck scales \cite{JennerMoreira}. 
Note that complex structures only can be built from interactions 
if open or closed orbits can  exist. This is
possible only for $n$ $>$ $1$.

Projeting ${\Pi}_{{{}_{e_n}}}$ and $ {\Pi}_{{{}_{m_p}}}$ 
in a space with three dimensions, i. e., for $n=3$, we have
\begin{equation}
\label{alfa8}
       {\Pi}_{{{}_{e_n}}} {}_{(n=3)}
       =  {{e}^2 \over {\hbar \; c}} ,
       \; \; \; \; \; \; \; \; 
       {\Pi}_{{{}_{m_p}}} {}_{(n=3)}  = 
        {{G \; {m_p}^2} \over {\hbar \; c}}  .
\end{equation}  
The above result shows that the
$\Pi$ numbers given by (\ref{alfa4}) and (\ref{alfa7}) 
are adimensional quantities obtained 
from a systematic use for interacting charges
of Vaschy-Buckingham Theorem, 
satisfying Laplace Equation in n-dimensions, that reduces, respectively,
to $\alpha$ and its gravitational analogue $\alpha_{{}_G}$ for $n$ $=$ $3$. 
It can be concluded that these $\Pi$s are in fact 
the generalized fine structure constants  
$\alpha_{{}_n}$ and ${\alpha_{}{{}_G}}_n$. 
This interpretation and the fact that
for $n$ $\neq$ 3 it is impossible to construct
an adimensional quantity with only $e_n$, $\hbar$ and $c$ 
suggest indeed that gravitation may play a relevant role
in atomic structure for $n$ $\neq$ 3.

Barrow and Tipler \cite{cap}
presented one among the infinite
possible adimensional combinations 
of the four fundamental n-dimensional
constants 
the number 
$\alpha_n^{\frac{(n-1)}{2}}$ 
$=$ ${\hbar}^{(2-n)} 
        {G_n}^{\frac{3-n}{2}}
        c^{(n-4)} {e_n}^{n-1} $
as the n-spatial
dimensionless constant of Nature. 
In this beautiful approach, which however does
not contemplate the Vaschy-Buckingham Theorem,
the singularity at $n$ $=$ $1$ disappears and
for $n$ = 1, 2, 3 and 4
each one of the fundamental
constants is absent.

In the interval $[0,1[$ the value of $\alpha_n$ begins 
at $10^{194}$ and diverges positively. Figure (1) shows the extreme
sensibility of $\alpha_n$ and ${\alpha_{}{{}_G}}_n$ 
on $n$ in the interval $]1,10]$. Only in a very short 
neighborhood of $n$ $=$ $3$, $\alpha_n$ $\sim$ 1 (for example
$\alpha_2$ $\sim$ $10^{-68}$,  $\alpha_4$ 
$\sim$  $10^{19}$ and $\alpha_n$ = 1 at n $\sim$ 3.067).
So, the error on the experimental value of $\alpha$
yields the narrow range in which we can find 
the dimensionality of the space in which we live.

\begin{figure}[htbp]
\begin{center}
\leavevmode
\epsfig{file=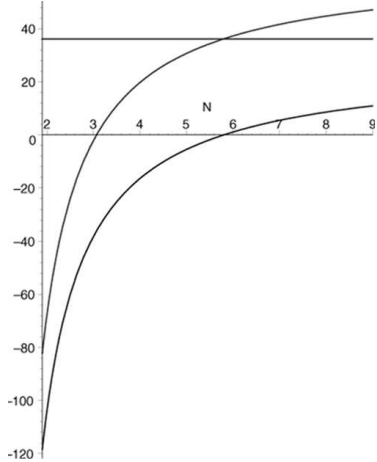,width=0.4\textwidth,angle=0}
\caption{{\small 
Dependence of log$_{10}[\alpha_n]$ (upper curve),
log${_{10}[\alpha_{}{{}_G}}_n]$ (lower curve) 
and log$_{10}[\alpha_n/{\alpha_{}{_G}}_n]$ 
(straight line) on n, the dimensionality of space.
The values of $\alpha_n $ and ${\alpha_{}{_G}}_n$
increase very radiply and tend to 
$\sim$ $10^{63}$ and $\sim$ $10^{27}$, respectively. }}
\end{center}
\end{figure}

In addition, 
the ratio between $\alpha_n$ and ${\alpha_{}{{}_G}}_n$
is a constant given by $e^2_n/(G_n m_p^2)$ =
 $e^2/(G m_p^2)$ 
$\sim$ $10^{36}$. 
This means that in  systems in which the balance between electric
and gravitational forces plays a major role - such as planets 
\cite{weiss},\cite{CarreRees} -
there are parameters - such as planetary radii - 
insensitive to dimensional change. This fact has deep astronomical, 
physical and biological implications. 
Note that, if the dimensionality of space
had evolved as a function of time, the fine 
tuning of electromagnetic interactions  
at atomic and molecular levels, necessary to
the existence of life, could not be maintained.
As a final remark, we note that the
relative standard uncertainty of $3.7 $ $\times$ $10^{-9}$ on the 
experimental value of $\alpha$ \cite{erro} implies a relative
standard uncertainty of 1.08 $10^{-13}$ on the three-dimensionality
of the physical space. 

\medskip

\noindent{\bf Acknowledgments}
The authors would like to thank F. Caruso for 
enlightening discussions on the early stages of this article. 
Financial assistance from Faperj (F. Casarejos) and CNPq (J.F.V. Rocha) 
is gratefully acknowledged.

\end{document}